%% file: main.tex
\documentclass{wlscirep}

\usepackage[utf8]{inputenc}
\usepackage[T1]{fontenc}
\usepackage{hyperref}
\usepackage{url}
\usepackage{booktabs}
\usepackage{amsfonts}
\usepackage{nicefrac}
\usepackage{microtype}
\usepackage{xcolor}
\usepackage{amsmath}
\usepackage{cleveref}
\usepackage{amsthm}

\usepackage{booktabs}
\usepackage{array}
\usepackage{xcolor} 
\usepackage{amsmath,amssymb,mathtools,bm}
\usepackage{bbding}
\allowdisplaybreaks

\usepackage{graphicx}
\usepackage{subcaption}

\newcommand{\name}{Archimedes}

\usepackage[normalem]{ulem}
\title{Scalable and Accurate Generative Sampling of Molecular Equilibrium Ensembles}

\author[1,2,$\#$,\Envelope]{Jiajun He}
\author[1,3,$\#$]{Denis Blessing}

\author[1,$\dagger$,\Envelope]{Yuanqi Du}
\author[1,$\dagger$,\Envelope]{Carles Domingo-Enrich}

\affil[1]{Microsoft Research New England, Cambridge, MA 02142, USA}
\affil[2]{University of Cambridge, Cambridge, CB2 1PZ, UK}
\affil[3]{Karlsruhe Institute of Technology, 76131 Karlsruhe, Germany}
\affil[$\#$]{Work done during internship at Microsoft Research.}
\affil[$\dagger$]{Equal supervision.}
\affil[\Envelope]{Correspondence. Email: jh2383@cam.ac.uk, yuanqidu@microsoft.com, carlesd@microsoft.com}

\begin{document}

\begin{abstract}
Sampling equilibrium ensembles is central to predicting molecular behavior and thermodynamic properties, yet becomes increasingly difficult as systems grow. Generative models offer a promising route to accelerating these calculations, but scaling them while retaining reliable statistical estimates remains a major challenge. Here we introduce \name{}, a multiscale diffusion framework that uses physical hierarchies to generate molecular configurations from coarse to fine while retaining evaluable path probabilities for statistical correction. The framework enables physically realistic all-atom protein generation beyond 1,000 atoms and supports highly accurate free energy calculations for solvent dielectric change, host-guest binding and protein mutation. 
\name{} also enables establishing transferable neural samplers that train primarily from molecular energies and transfer to unseen peptide sequences.
Beyond biomolecules, the spatial hierarchy transfers to CO$_2$ boxes containing eight times as many molecules as in training and reproduces both the local structure of molten and glassy Cu-Zr and the reduced free energy difference between them.
Together, these results establish physical hierarchies as a route to scaling generative sampling while preserving access to molecular thermodynamics.
\end{abstract}

\maketitle

\section{Introduction}

Generative modeling of equilibrium distributions has advanced substantially in recent years, from Boltzmann generators to biomolecular emulators such as BioEmu.\cite{noe2019boltzmann,lee2026predicting,schebek2026scalable,lewis2025bioemu} Yet reconciling thermodynamic accuracy with computational scalability remains challenging. Methods that support statistical correction to a target Boltzmann distribution and rigorous free energy estimation face substantial computational barriers as system size grows.\cite{peng2025perturbation,schebek2026assessing} BioEmu demonstrates that generative models can reproduce conformational diversity and thermodynamic properties across a broad range of proteins,\cite{lewis2025bioemu} but such predictive accuracy does not by itself establish sampling from the Boltzmann distribution of a specified all-atom energy function. Combining scalable all-atom generation with explicit thermodynamic correction remains an important goal for molecular simulation.

This gap has broad implications for molecular simulation. Thermodynamically reliable all-atom generation would allow generative models to support free energy calculations of binding affinities and mutation-induced changes in protein stability.\cite{noe2019boltzmann,wirnsberger2020targeted} The same physical fidelity is also necessary to learn neural samplers directly from energy functions, reducing reliance on costly or even inaccessible simulation datasets.\cite{havens2026flowsampling,blessing2026bridge}  Both opportunities depend on generating physically realistic configurations while capturing the diversity of molecular conformations, a requirement that becomes increasingly demanding as system size grows.

Here, we introduce \name{}, a multiscale diffusion model for all-atom equilibrium sampling and free energy estimation. The model exploits the separation between global conformation and local geometry: it first generates coarse structural information, then resolves finer degrees of freedom conditioned on the structure already present. 
The hierarchy follows the physical organization of each system. For proteins and molecular complexes, selected atoms describe the overall arrangement and anchor the generation of local atomic detail. In bulk systems, local particle counts describe the coarse density distribution, followed by molecular positions. This factorization separates global variation from shorter local displacements while retaining computable weights for free energy estimation.

\name{} scales accurate all-atom protein generation exceeding 1,000 atoms and bulk systems with more than 5,000 atoms, with comparisons showing that the multiscale hierarchy improves both sample quality and thermodynamic estimation relative to single-level constructions. 
The generation process also carries trackable path probabilities, enabling estimation of free energy differences with nonequilibrium approaches, including changes in solvent conditions, relative host--guest binding affinities, and a mutation-induced stability change.
The framework also supports amortized neural samplers learned primarily from energy evaluations, with limited reference data, and generates peptides with sequences and lengths absent from training. 
For bulk systems, a spatial occupancy hierarchy improves generation and free energy estimation and supports transfer from smaller training systems to larger systems.

 \begin{figure}[!t]
\centering
\includegraphics[width=\textwidth]{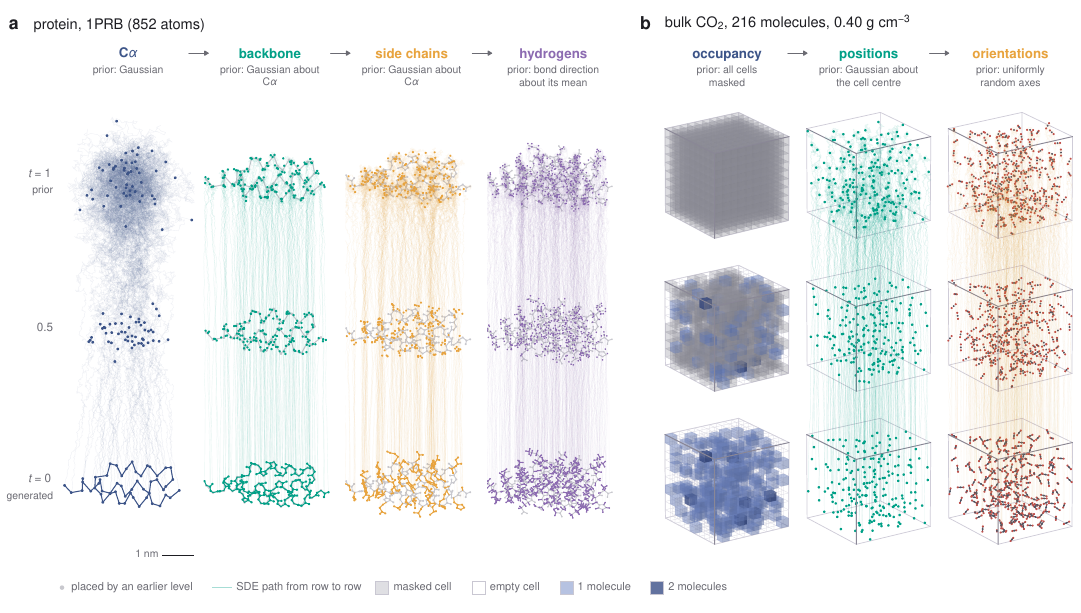}
\caption{\textbf{Level-by-level generation with \name{}.}
Columns show successive hierarchy levels from coarse to fine. Rows show the prior ($t=1$), an intermediate state ($t=0.5$) and the generated configuration ($t=0$), from top to bottom. Thin lines show denoising trajectories.
\textbf{a}, Generation of 1PRB through four levels: C$\alpha$ atoms, remaining backbone heavy atoms, side-chain heavy atoms and hydrogens. Atoms generated at coarser levels are grey and remain fixed. All stages use the same view and scale.
\textbf{b}, Generation of a box of 216 CO$_2$ molecules through cell occupancies, molecular positions and orientations. Masked cells are grey; revealed occupied cells are light blue for one molecule and dark blue for two. Empty cells are outlined. Molecular positions are generated relative to the assigned cell centres, followed by orientations.}
\label{fig:hierarchy}
\end{figure}

\section{Results}

\subsection*{Overview of \name{}}

\name{} decomposes all-atom generation into a hierarchy of conditional diffusion models. It first generates coarse variables, then resolves finer degrees of freedom conditioned on the structure already present. Writing these variables as $y_1,\ldots,y_L$, the generative distribution factorizes as
\begin{equation}
p_\theta(y_1,\ldots,y_L)
=
\prod_{\ell=1}^{L}
p_{\theta,\ell}(y_\ell\mid y_{<\ell}),
\label{eq:hierarchy-factorization}
\end{equation}
where $y_{<\ell}$ denotes the coarser variables. For proteins and molecular complexes, selected atoms define the coarse structure, and finer atoms are generated relative to their positions (Fig.~\ref{fig:hierarchy}a). For bulk systems under periodic boundary conditions, local particle counts describe the coarse spatial distribution, followed by molecular positions and orientations (Fig.~\ref{fig:hierarchy}b).

Each level uses continuous or discrete diffusion. For continuous levels fitted to molecular configurations, the forward noising process has closed-form Gaussian transitions and reaches its Gaussian prior exactly. A neural network parameterizes Gaussian denoising transitions conditioned on the coarser structure. At the occupancy level, masked discrete diffusion generates cell counts through categorical transitions. The noising and denoising transition densities, together with the priors and molecular energy, provide path probabilities for correcting generated ensembles and estimating free energy (Methods).

\subsection*{\name{} enables accurate all-atom modeling of protein Boltzmann distributions}

\begin{figure}[!t]
  \centering
  \includegraphics[width=\textwidth]{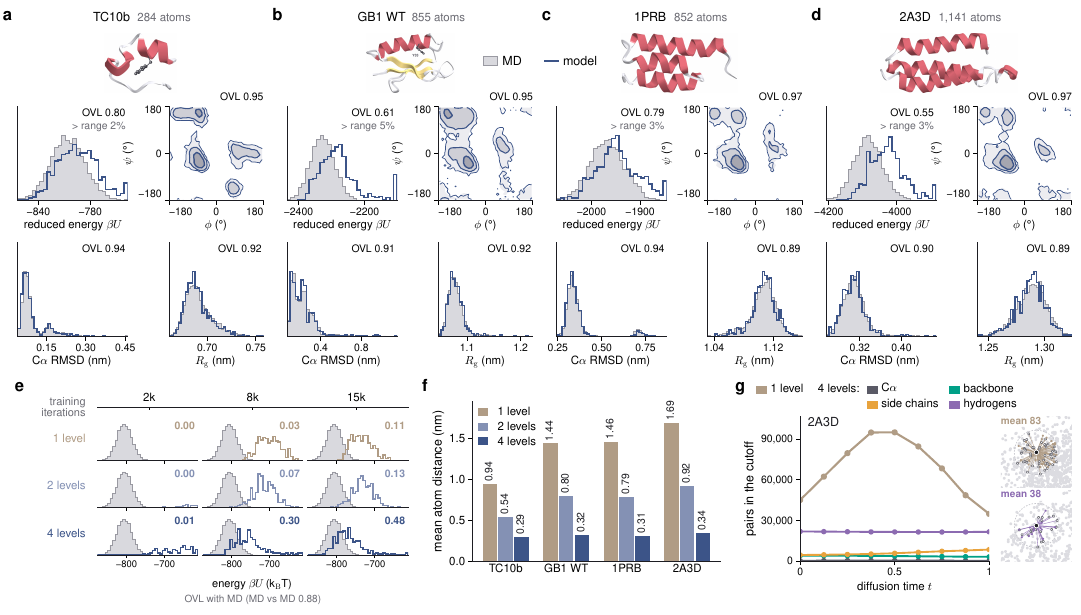}
\caption{\textbf{\name{} enables accurate all-atom protein sampling.}
\textbf{a-d}, TC10b, GB1 wild type, 1PRB and 2A3D. Generated ensembles (blue) and reference molecular dynamics (MD, grey) are compared through reduced potential energy $\beta U$, backbone dihedrals $(\phi,\psi)$, C$\alpha$ root-mean-square deviation (RMSD) and radius of gyration $R_\mathrm{g}$. OVL denotes distribution overlap; percentages indicate samples outside the displayed energy range.
\textbf{e}, Energy distributions during training for one-, two- and four-level models on TC10b. The two-level model generates heavy atoms before hydrogens. Numbers give overlap with MD.
\textbf{f}, Mean atomwise distance between independently drawn prior and training configurations, with prior widths matched to the data at each level.
\textbf{g}, Atom pairs within the $0.5\,\mathrm{nm}$ cutoff during diffusion in 2A3D. 
Insets show neighbors of a hydrogen at $t=0.5$ for single-level generation (top) and the hydrogen level (bottom). Filled and open circles denote hydrogens and heavy atoms; dashed circles mark the cutoff. Inset labels give the mean number of neighbours within the cutoff.}
  \label{fig:four-proteins}
\end{figure}

 A significant challenge in modeling protein equilibrium ensembles at all-atom resolution is the rapidly growing complexity of the underlying distribution with system size. 
 Even when solvent is treated implicitly, explicitly including all heavy atoms and hydrogen atoms adds many tightly constrained degrees of freedom that must be sampled together with collective structural changes. 
 The demands are even greater for free energy estimation across solvent conditions, ligand structures or protein mutations, which requires both access to the relevant conformations and accurate accounting of their statistical weights.

\name{} captures conformational variation while generating physically realistic all-atom structures across TC10b, GB1 wild type, 1PRB and 2A3D, spanning 284 to 1,141 atoms in implicit solvent. The generated ensembles capture the main backbone dihedral basins observed in reference molecular dynamics (MD) and reproduce the distributions of C$\alpha$ RMSD and radius of gyration (Fig.~\ref{fig:four-proteins}a-d). Beyond this conformational agreement, their potential energy distributions also overlap substantially with MD. Energy agreement provides a stringent test of atomic geometry: distorted bonds or steric clashes can make a configuration energetically unfavourable even when its overall conformation appears plausible. The model therefore reproduces protein ensemble properties while maintaining realistic local geometry at the thousand-atom scale.

Without \name{}, resolving these structural scales jointly remains difficult even for TC10b, the smallest protein considered here. We compared \name{} with one- and two-level baselines using matched parameter counts and optimization settings. Both baselines adapt BioEmu's joint generation of backbone conformations to our all-atom setting.\cite{lewis2025bioemu} The one-level model generates all atoms jointly, whereas the two-level model generates heavy atoms jointly and then hydrogens conditioned on their positions. At 15,000 training iterations, the four-level model achieves approximately four times the energy-distribution overlap with MD of either baseline (Fig.~\ref{fig:four-proteins}e).

The hierarchy keeps most atomic displacements local and reduces the number of neighbouring atom pairs that each network needs to consider.
In single-level generation, the spatial scale grows with the overall extent of the protein. The hierarchy concentrates this size dependence at the coarse level, while most atoms are generated at finer levels on local geometric scales that change little across proteins (Fig.~\ref{fig:four-proteins}f).
The hierarchy also limits the growth of the size of molecular graph during diffusion. Noising all atoms together creates many additional neighbours at intermediate times. In the hierarchy, each network instead processes the atoms at its own level together with the fixed coarser structure, keeping the graphs smaller and their sizes more stable throughout diffusion (Fig.~\ref{fig:four-proteins}g).

\subsection*{\name{} enables efficient free energy estimation for biomolecules}

\begin{figure}[!t]
  \centering
  \includegraphics[width=\textwidth]{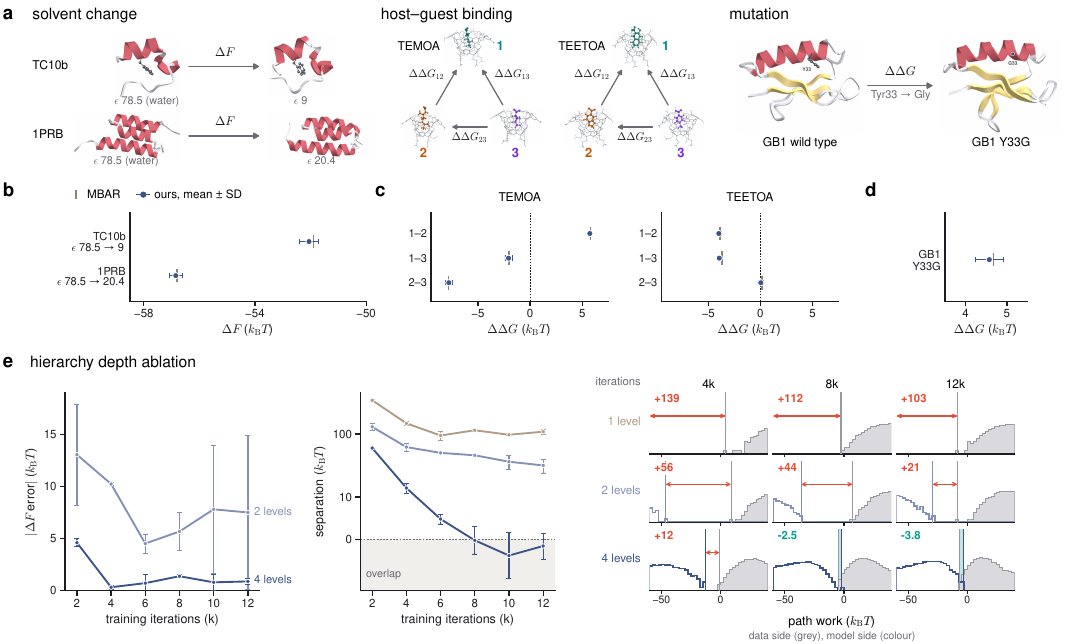}
\caption{\textbf{\name{} accurately estimates biomolecular free energy.}
\textbf{a}, Molecular structures for solvent changes, host-guest binding and protein mutation.
\textbf{b}, Free energy differences between solvent dielectrics, $\Delta F=F(78.5)-F(\epsilon_{\mathrm{low}})$, with $\epsilon_{\mathrm{low}}=9$ for TC10b and $20.4$ for 1PRB.
\textbf{c}, Relative binding free energy for three guest pairs in each of TEMOA and TEETOA, with $\Delta\Delta G_{ij}=\Delta G_i-\Delta G_j$.
\textbf{d}, Mutation stability change for GB1 Y33G.
In \textbf{b-d}, grey ticks denote MBAR references; blue points and error bars show \name{} means and standard deviations over three evaluations.
\textbf{e}, Effect of hierarchy depth on TC10b at $\epsilon=78.5$. Left, absolute error in $\Delta F$ relative to MBAR for two- and four-level models, each paired with the same final model at $\epsilon=9$. Middle, sampled work separation, $\min(-w_f)-\max(w_b)$, for one, two and four levels. The left and middle panels show means and standard deviations over two disjoint groups of paths.
Right, work distributions at selected training checkpoints: forward noising paths from MD ($-w_f$, grey) and reverse generative paths ($w_b$, colour). Annotations give the signed separation; negative values indicate overlap (green), and positive values indicate a gap (red).}
  \label{fig:free energy}
\end{figure}

Accurate molecular generation enables \name{} to predict thermodynamic preferences through free energy differences. 
These quantify solvent effects, relative binding affinities and mutation stability (Fig.~\ref{fig:free energy}a). Building on nonequilibrium thermodynamics, the learned diffusion processes provide free energy estimates from molecular energy and path probabilities.\cite{jarzynski1997,vaikuntanathan2008escorted,crooks1999entropy} We combine information from the noising and generative paths using Bennett's acceptance ratio (BAR), and obtain free energy differences from the endpoint estimates (Methods).\cite{bennett1976}

Accurate equilibrium ensembles alone do not ensure precise free energy estimates immediately, because the calculation depends on the entire diffusion trajectory. Overlap between the forward and reverse work distributions reflects agreement between the two directions and affects the precision of the estimate.\cite{hahn2010convergence} A common way to improve generated structures is to relax them after sampling. However, relaxation that satisfies detailed balance with the target leaves the work unchanged when included consistently in the calculation, and therefore cannot restore this overlap (\hyperref[eq:si-relaxation]{Supplementary Eq.~(\ref*{eq:si-relaxation})}). Free energy estimation thus places a stronger demand on the learned diffusion process than the quality of its generated configurations alone.

\name{} accurately recovers the thermodynamic effects of solvent changes, host-guest binding and protein mutation in close agreement with multistate Bennett acceptance ratio (MBAR) references (Fig.~\ref{fig:free energy}b-d). This agreement extends to subtle differences in binding preference: the mean predictions recover the guest ranking in both TEMOA and TEETOA, including the TEETOA guests with nearly equal affinities. The binding and mutation calculations each combine four independently trained models, showing that the accuracy extends to complete thermodynamic cycles. For GB1 Y33G, \name{} recovers the mutation-induced stability change directly from the physical molecules at each endpoint. It thus reproduces the alchemical benchmark while removing the need to construct a shared topology with ghost atoms or sample intermediate alchemical states.

The hierarchy contributes substantially to this thermodynamic accuracy (Fig.~\ref{fig:free energy}e). On TC10b, the four-level model establishes and maintains overlap between the sampled forward and reverse work ranges during training, whereas the one- and two-level models remain largely separated. The corresponding solvent free energy estimates have lower error with four levels consistently across training iterations, with reductions approaching an order of magnitude relative to the two-level model. To reduce the remaining fluctuations along the diffusion paths, we further introduce \emph{conditional path averaging}, which averages contributions from alternative trajectories while keeping their endpoints fixed. On GB1, this improves agreement between the forward and reverse free energy estimates at the same network evaluation budget, providing an additional means of improving estimation beyond the hierarchy itself (Methods and \hyperref[fig:path-averaging]{Supplementary Fig.~\ref*{fig:path-averaging}}).
Together, \name{} brings diffusion-based free energy estimation to all-atom proteins and molecular complexes, with the accuracy needed to predict binding preferences and mutation-induced changes in protein stability.

\begin{figure}[!t]
  \centering
  \includegraphics[width=\textwidth]{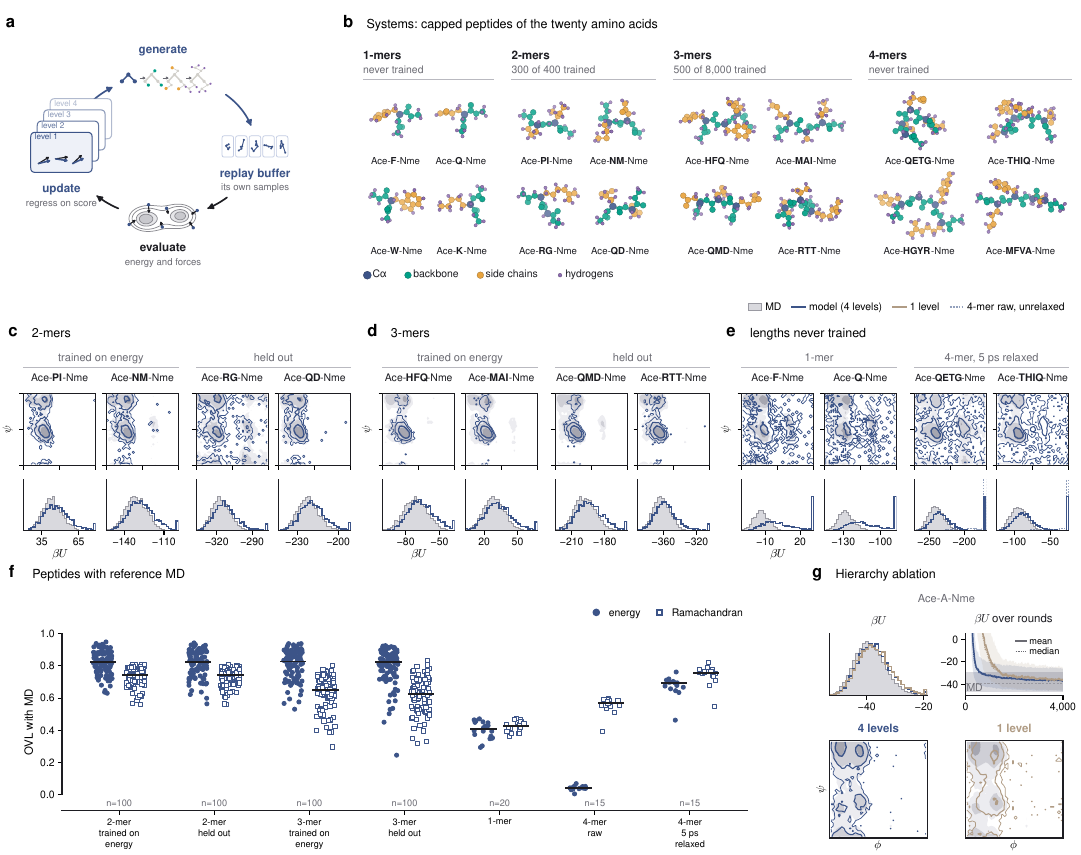}
  \caption{\textbf{Learning transferable peptide samplers from energy.}
  \textbf{a}, Overview of learning from energy. Generated configurations enter a replay buffer, and their energy and forces guide updates to the diffusion models at all hierarchy levels.
  \textbf{b}, Peptide systems and training split. Lengths exclude terminal caps; atoms are coloured by hierarchy level. Shared training includes reference MD configurations for 40 homopeptides.
  In \textbf{c}-\textbf{g}, grey denotes MD, blue the four-level model and brown the one-level model.
  \textbf{c,d}, Generated and reference distributions for training and unseen two-residue (\textbf{c}) and three-residue (\textbf{d}) sequences. Top, backbone Ramachandran densities pooled over residues; bottom, reduced potential energy $\beta U$.
  \textbf{e}, Transfer to unseen chain lengths. One-residue samples are shown as generated and four-residue samples after 5\,ps of Langevin relaxation. Dotted lines show unrelaxed energy distributions.
  \textbf{f}, Energy (filled circles) and Ramachandran (open squares) overlap with MD. Points represent individual sequences, bars show group medians and $n$ gives group sizes. MD-supervised sequences are excluded.
  \textbf{g}, Effect of hierarchy depth on separately trained alanine dipeptide samplers: energy and Ramachandran distributions after 4,000 rounds, and energy evolution during training. Curves show the mean (solid) and median (dotted); shading indicates the 5th-95th percentile range. The MD mean and corresponding range are shown in grey.}
  \label{fig:peptides}
\end{figure}

\subsection*{\name{} learns transferable molecular samplers from energy functions}

Obtaining reference ensembles and training a separate model for every molecule would limit the reach of generative sampling. Previous work has shown that a single generative model trained on peptide simulations can sample unseen sequences.\cite{klein2024transferable,tan2025amortized}
\name{}, instead, addresses both costs by directly learning primarily from molecular energy functions and sharing the sampler across molecules. 
Training alternates between generating configurations and using their energy and forces to update the model (Fig.~\ref{fig:peptides}a).\cite{havens2026flowsampling,blessing2026bridge,cheng2026atlas} Each complete configuration supplies a training signal to every hierarchy level, coupling the learning of global conformation and local geometry through the same molecular energy function. Conditioning on atom identities and bond connectivity allows the trained sampler to generate configurations for different molecules (Methods).

We train a shared sampler across 800 capped peptide sequences containing two or three residues, supplemented by reference MD configurations for 40 homopeptides within this pool. 
The total force evaluation budget, even including the reference MD configurations, corresponds to only approximately $2.5\,\mathrm{ns}$ of reference MD per training sequence on average. 
For both two- and three-residue peptides, \name{} reproduces the reference energy distributions, with comparable agreement on training and unseen sequences in both energy and backbone dihedrals (Fig.~\ref{fig:peptides}c,d,f). The close agreement in energy supports the physical realism of the generated all-atom structures, although an extended backbone basin remains underrepresented in the three-residue ensembles. The comparable performance across training and unseen sequences demonstrates that the learned sampling behaviour transfers to new amino acid combinations without sequence-specific reference configurations or further training.
Beyond sequence transfer, \name{} also supports transfer to chain lengths absent from training. The same trained model generates capped peptides containing one or four residues, extending beyond the two- and three-residue sequences used for learning (Fig.~\ref{fig:peptides}e,f). For four-residue peptides, brief relaxation improves agreement with the reference energy distributions.

A separate comparison on alanine dipeptide examines learning from energy alone with one- and four-level samplers (Fig.~\ref{fig:peptides}g). The four-level sampler reaches low-energy configurations earlier and reproduces the backbone conformational distribution more accurately. 
Together with transfer across unseen sequences, these results demonstrate how \name{} supports learning reusable molecular samplers while significantly reducing dependence on molecule-specific reference simulations.

\begin{figure}[!t]
  \centering
  \includegraphics[width=\textwidth]{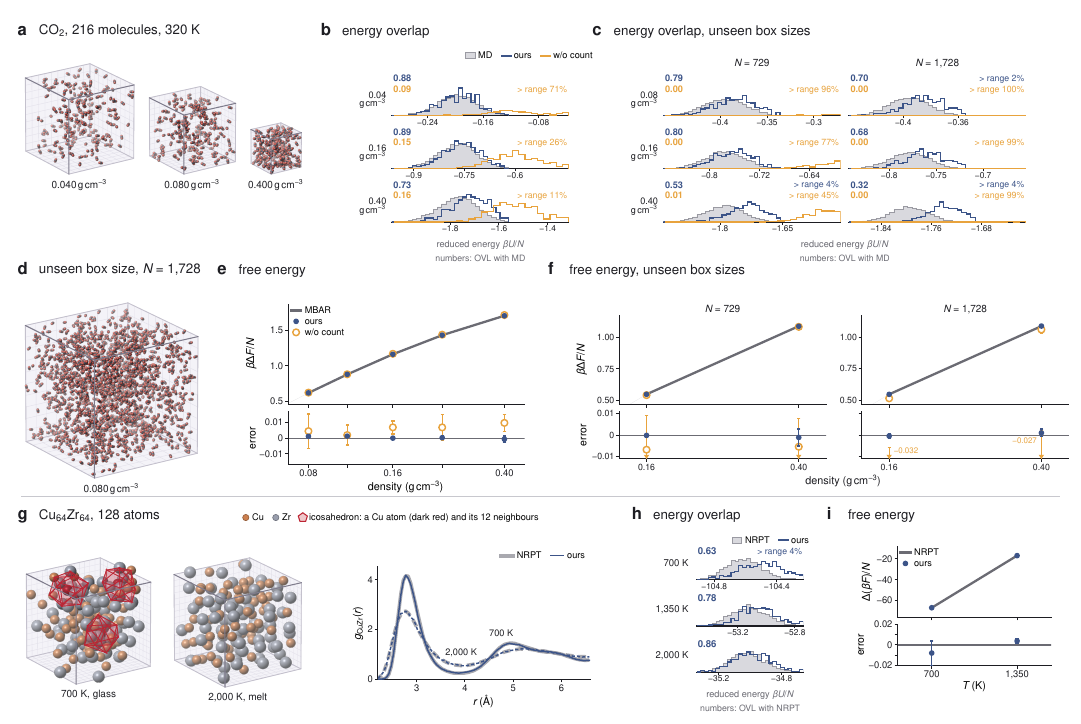}
\caption{\textbf{Sampling and free energy of bulk CO$_2$ and a glass-forming Cu-Zr alloy.}
Blue denotes \name{}, orange the model without cell occupancies and grey the reference.
\textbf{a,d}, Boxes of 216 and 1,728 CO$_2$ molecules at $320\,\mathrm{K}$, shown at the same scale.
\textbf{b,c}, Distributions of reduced potential energy per molecule, $\beta U/N$, compared with molecular dynamics at $N=216$ (\textbf{b}) and after transfer to $N=729$ and $1{,}728$ (\textbf{c}).
\textbf{e,f}, Compression free energy per molecule, $\beta\Delta F/N$, relative to $0.040\,\mathrm{g\,cm^{-3}}$ for $N=216$ (\textbf{e}) and $0.080\,\mathrm{g\,cm^{-3}}$ for the larger boxes (\textbf{f}).
\textbf{g}, Cu$_{64}$Zr$_{64}$ representative configurations. Polyhedra mark Cu-centred icosahedra. Solid curves show  $700\,\mathrm{K}$ and dashed curves show $2{,}000\,\mathrm{K}$.
\textbf{h}, Reduced potential energy per atom compared with NRPT at three temperatures.
\textbf{i}, Change in reduced free energy per atom, $\Delta(\beta F)/N$, relative to $2{,}000\,\mathrm{K}$.
Histogram labels give overlap with the reference and the percentage of generated samples above the plotted range. Free energy points and error bars show BAR means and standard deviations over four paired evaluations. Lower panels show deviations from MBAR for CO$_2$ and from the NRPT stepping-stone reference for Cu-Zr.
}
  \label{fig:fluid}
\end{figure}

\subsection*{\name{} extends hierarchical sampling and free energy estimation to bulk systems}

In proteins, covalent structure guides the placement of atoms relative to one another. In bulk systems, molecules are mostly permutation invariant, so \name{} builds its hierarchy from the spatial distribution of particles. 
Discrete diffusion first generates the number of particles in each cell. 
Continuous diffusion then places the particles relative to their assigned cell centers. 
For rigid molecules such as CO$_2$, a further level generates orientations conditioned on the positions (Fig.~\ref{fig:hierarchy}b).

\name{} reproduces the MD potential energy distributions of 216 rigid TraPPE CO$_2$ molecules at $320\,\mathrm{K}$ across six densities from $0.040$ to $0.400\,\mathrm{g\,cm^{-3}}$ (Fig.~\ref{fig:fluid}a,b). 
For all five compressions from the lowest density, its estimates of the Helmholtz free energy change per molecule agree with MBAR to within $0.2\%$ (Fig.~\ref{fig:fluid}e). To assess the contribution of cell occupancies, we compared \name{} with a two-level torus model that generates positions over the entire periodic box by wrapped diffusion and then generates orientations. 
This model shows little energy overlap with MD. 
The occupancy level reduces the overall error in the five compression free energy by nearly an order of magnitude (Fig.~\ref{fig:fluid}b,e).

\name{} also transfers to larger CO$_2$ systems without retraining. The occupancy CTMC uses the same local convolutional kernels to predict each cell's count from neighboring cells, so it can generate occupancies on a larger lattice. At fixed density, we enlarge the lattice by adding cells of comparable width; hence, the position model still places each molecule by sampling a short displacement from its assigned cell center.
A model trained on 216 molecules transfers to boxes containing up to 1,728 molecules, an eightfold increase in system size (Fig.~\ref{fig:fluid}c,d).
At every size tested, the estimated free energy changes for compression agree with MBAR to within $0.5\%$. 
In the largest box, the occupancy hierarchy reduces the errors for these two compressions from $0.080$ to $0.160$ and $0.400\,\mathrm{g\,cm^{-3}}$ by factors of up to two orders of magnitude, relative to the model without the occupancy hierarchy (Fig.~\ref{fig:fluid}f).

\name{} also enables sampling and free energy estimation in glass-forming materials. Extending the spatial hierarchy to a metallic glass former requires the model to generate distinct amorphous structures while reproducing the local atomic order that develops as the melt cools.\cite{sheng2006packing,ding2015packing}
We model Cu$_{64}$Zr$_{64}$ with the MACE-MPA-0 interatomic potential,\cite{batatia2025foundation} using a single \name{} model to describe the melt at $1{,}350$ and $2{,}000\,\mathrm{K}$ and the glass at $700\,\mathrm{K}$.
Our data and reference are collected with non-reversible parallel
tempering (NRPT) \cite{syed2022non}.
The generated ensembles reproduce the splitting of the second peak in the Cu-Zr radial distribution function on cooling from liquid to glassy states (Fig.~\ref{fig:fluid}g). 
Even in the glass, the generated potential energy distribution has an overlap of $0.63$ with NRPT (Fig.~\ref{fig:fluid}h). At both lower temperatures, the predicted change in reduced free energy relative to the $2{,}000\,\mathrm{K}$ melt agrees with the NRPT reference to within $0.01$ per atom (Fig.~\ref{fig:fluid}i).

\section{Discussion}

Here we introduce \name{}, a multiscale diffusion framework for generating molecular equilibrium ensembles and estimating their free energy. Organizing generation through a physical hierarchy improves the ability of diffusion models to capture conformational variation while preserving realistic atomic geometry. Earlier work explored coarse-to-fine generation for peptides and for reconstructing all-atom conformations from coarse molecular structures.\cite{abdin2024pepflow,wang2022backmapping} In \name{}, molecular structure and spatial occupancies provide different forms of coarse information, allowing the same principle to apply to proteins, molecular complexes and periodic bulk systems. This decomposition addresses a central scaling challenge by allowing global organization and local atomic detail to be learned at their respective scales.
Using hierarchical diffusion, \name{} for the first time achieves both accurate sampling of Boltzmann ensembles and accurate estimation of free energy differences from path weights in all-atom proteins at the thousand-atom scale and in disordered bulk systems extending to more than 5,000 atoms.

The benefits extend beyond reproducing reference ensembles. Improved agreement between forward and reverse diffusion paths supports quantitative free energy calculations, while conditional path averaging reduces fluctuations along those paths. 
Together, these developments allow thermodynamic preferences to be recovered from models of the endpoints directly, including comparisons between molecules with different topologies and atom counts, without intermediate alchemical states or ghost atoms. The hierarchy also supports learning primarily from molecular energy and sharing the resulting sampler across molecules. Transfer across sequences and system sizes shows that learned sampling behaviour can be reused beyond the systems used for training, reducing the need to obtain a separate reference ensemble and fit a separate model for every target.

Several limitations and future improvements remain. Free energy estimation still requires representative equilibrium configurations at each endpoint and sufficient overlap between forward and reverse paths. Learning from energy could reduce dependence on reference configurations, but it remains susceptible to missing conformational modes despite reproducing the energy distribution. 
Work overlap also deteriorates in larger, denser bulk systems, motivating training objectives that directly improve agreement between forward and reverse paths. Addressing these challenges would support broader amortization across molecules and thermodynamic states.
Thermodynamic predictions also inherit errors in the underlying energy function, and the biomolecular calculations here use implicit solvent. Future work could combine molecular and spatial hierarchies to describe complex solutes surrounded by explicit water.

\section{Methods}

At each hierarchy level, the forward noising process adds noise to continuous coordinates or masks cell counts, and the reverse process generates these variables from the corresponding prior. Continuous diffusion time runs from $t=0$ at the molecular configuration to $t=1$ at the prior; generation proceeds from $t=1$ to $t=0$.

\subsection*{Hierarchical decomposition}

We represent a molecular configuration $x$ using variables $y_1,\ldots,y_L$, ordered from coarse to fine. Following the conditional factorization in Eq.~\eqref{eq:hierarchy-factorization}, each level generates its variables with the coarser variables held fixed.

For proteins, the four levels generate the C$\alpha$ trace, the remaining backbone heavy atoms, the side-chain heavy atoms and the hydrogens. We centre the C$\alpha$ coordinates to remove overall translation. Backbone and side-chain coordinates are expressed as displacements from the C$\alpha$ atom of the same residue, and hydrogen coordinates as displacements from their bonded heavy atoms. For configurations with constrained X-H bond lengths, the hydrogen level generates only the bond directions. For host-guest complexes, selected atoms on both molecules are generated through two coarse levels. Subsequent levels generate the remaining heavy atoms relative to their assigned coarse atoms, followed by hydrogens relative to their bonded heavy atoms (\hyperref[fig:hg-levels]{Supplementary Fig.~\ref*{fig:hg-levels}}).

For bulk systems under periodic boundary conditions, the coarse variables are the cell counts on a spatial lattice. Let $R_i$ denote the position of particle $i$ or the centre of molecule $i$, and let $c_s$ denote the centre of cell $s$. Given a configuration $x$, we sample a cell label $a_i$ with probability
\begin{equation}
r(a_i=s\mid x)
=
\frac{
\exp\!\left[-d_{\mathrm{PBC}}(R_i,c_s)^2/(2\sigma_{\mathrm a}^2)\right]
}{
\sum_{s'}
\exp\!\left[-d_{\mathrm{PBC}}(R_i,c_{s'})^2/(2\sigma_{\mathrm a}^2)\right]
},
\end{equation}
where $d_{\mathrm{PBC}}$ is the minimum-image distance and $\sigma_{\mathrm a}$ controls the spatial width of the assignment. Labels are sampled independently conditional on $x$, giving
$r(a\mid x)=\prod_i r(a_i\mid x)$
and cell counts
$n_s=\sum_i\mathbf{1}[a_i=s]$.

During generation, the occupancy level first generates the cell counts. Continuous levels then generate positions relative to the assigned cell centres. For rigid CO$_2$, a further level generates molecular orientations conditional on these positions. The assignments are probabilistic, so a cell label does not restrict a molecule to remain within that cell.

Including the cell labels defines an augmented target with energy
\begin{equation}
U_{\mathrm{aug}}(x,a)
=
U(x)-\beta^{-1}\log r(a\mid x),
\qquad
\beta=(k_{\mathrm B}T)^{-1},
\end{equation}
where $U(x)$ is the molecular potential energy. The corresponding Boltzmann weight is
$e^{-\beta U_{\mathrm{aug}}(x,a)}
=e^{-\beta U(x)}r(a\mid x)$.
Because $\sum_a r(a\mid x)=1$, summing over cell labels recovers the original Boltzmann weight and leaves the partition function unchanged.

\subsection*{Diffusion within each level}

Each hierarchy level uses a continuous or discrete diffusion model, with the coarser variables held fixed. 
For continuous variables, the forward process adds noise to the coordinates, and the reverse process generates coordinates from a Gaussian prior.\cite{ho2020ddpm,song2021sde}
For discrete cell counts, the forward process masks their values, and the reverse process generates them from a fully masked lattice through categorical transitions. 
Both constructions have evaluable transition probabilities, which, together with the priors, determine the probability of each path.

For the continuous models for protein and molecular complex, we fit the reverse process to reference configurations. 
The forward process is a noising process with closed-form Gaussian transitions that reaches its Gaussian prior exactly. 
The network predicts the clean coordinates and their uncertainty, which determine the mean and covariance of each Gaussian reverse transition (\hyperref[eq:si-gaussian-reverse]{Supplementary Eq.~(\ref*{eq:si-gaussian-reverse})}). 
Training maximizes the path-space evidence lower bound on the reference configurations.\cite{kingma2021vdm}

For bulk systems, the continuous position and orientation levels use the Gaussian transitions described above. 
The discrete occupancy level uses masked diffusion, implemented as a continuous-time Markov chain (CTMC).\cite{campbell2022discrete} Its reverse process reveals cell counts one cell at a time in a uniformly sampled order.
Each categorical transition is constrained by the number of particles or molecules remaining to be assigned. We train the model by cross-entropy on cell counts sampled from MD configurations.

For learning from energy, the network instead directly predicts the drift $u_\theta$ of the reverse process, following Flow Sampling.\cite{havens2026flowsampling} Denoting the continuous coordinates by $z_t$ and the normalized prior density by $\rho$, the generative dynamics are
\begin{equation}
\mathrm d z_t
=
u_\theta(z_t,t)\,\mathrm dt
+
\sqrt{2\lambda(1-t)}\,\mathrm d\bar W_t,
\qquad
z_1\sim\rho,
\end{equation}
integrated from $t=1$ to $t=0$. Here $\bar W_t$ is reverse-time Brownian motion and $\lambda>0$ controls the diffusion strength. The level index and conditioning on the coarser variables are omitted. Sampling and path evaluation use the same discretized transitions. The drift training objective is described below.

\subsection*{Nonequilibrium free energy estimation}

We estimate the configurational partition function $Z$ of each thermodynamic state from diffusion paths through the hierarchy. Let $\Gamma$ denote a complete path, including all continuous and discrete steps but excluding $x$, which is its configuration at $t=0$. Starting from configuration $x$, the forward noising process samples the rest of the path with conditional density $Q(\Gamma\mid x)$. 
Starting from the normalized priors, the reverse process samples the entire path, including $x$, with joint density $P_\theta(\Gamma, x)$.
The dimensionless work is the log importance weight
\begin{equation}
w(\Gamma,x)
=
-\beta U(x)
+
\log\frac{Q(\Gamma\mid x)}{P_\theta(\Gamma, x)}.
\end{equation}
Averaging the importance weights $e^w$ gives the partition function and hence the free energy:
\begin{equation}
Z=\mathbb E\!\left[e^w\right],
\qquad
F=-\beta^{-1}\log Z.
\end{equation}
This is the importance-weight form of the escorted Jarzynski equality, with forward and reverse work distributions related by the Crooks fluctuation theorem.\cite{jarzynski1997,vaikuntanathan2008escorted,crooks1999entropy,he2025feat}
In practice, Bennett's acceptance ratio (BAR) combines work values from generative paths and paths obtained by noising equilibrium configurations to estimate $\log Z$.\cite{bennett1976} Under this convention, $w=-w_f$ for forward paths and $w=w_b$ for reverse paths, matching the quantities plotted in the figures.

\paragraph{Thermodynamic cycles}
We fitted a separate \name{} model to each endpoint and estimated its free energy from forward and reverse diffusion paths using BAR (Fig.~\ref{fig:free energy}b-d). For the references, we applied MBAR to equilibrium samples from all states of each NRPT ladder.
For TC10b and 1PRB, we calculated the free energy change on increasing the solvent dielectric from 9 (TC10b) or 20.42 (1PRB) to 78.5.
For each guest in TEMOA and TEETOA, we added the standard-state restraint correction to the coupling free energy $F^{\mathrm{coupled}}-F^{\mathrm{decoupled}}$ to obtain its standard binding free energy $\Delta G_i$.\cite{gallicchio2010bedam} We report $\Delta\Delta G_{ij}=\Delta G_i-\Delta G_j$. The correction cancels for all TEMOA guest pairs. In TEETOA, G4 has a smaller restraint volume than G1 and G5, so only pairs involving G4 require a correction (\hyperref[sec:si-binding-offset]{Supplementary Section~\ref*{sec:si-binding-offset}}).
For GB1 Y33G, the change in folding free energy was the Tyr-to-Gly mutation free energy in folded GB1 minus that in the capped tripeptide Gln-Tyr-Ala, representing the unfolded state.\cite{koenekoop2025mutation}
The MBAR reference used intermediate states containing atoms from both residues. 
Atoms absent from a physical endpoint remained as ghost atoms with their nonbonded interactions with the surroundings switched off. \name{} used the four physical endpoints directly without any ghost atoms, each with its own topology.

\subsection*{Conditional path averaging}

Two diffusion paths can connect the same molecular configuration $x$ to the same prior draw and still have different work values because their intermediate states differ. We introduce conditional path averaging to reduce this source of variation, in a similar way as Ref.~\citenum{elvira2019generalized} We divide each Gaussian diffusion chain into segments and hold the two endpoints of each segment fixed. We retain the original trajectory between those endpoints as candidate $m=1$ and draw $M-1$ alternatives from the forward noising process conditioned on the same endpoints. This conditional distribution is a Gaussian bridge that we can sample exactly.

For each candidate, we calculate how likely its intermediate steps are under the reverse and forward processes. Let $q$ and $p_\theta$ denote the forward and reverse transition densities, respectively. For segment $k$, spanning steps $a_k$ to $b_k$, the ratio for candidate $m$ and its average over all $M$ candidates are
\begin{equation}
R_{km}
=
\prod_{j=a_k+1}^{b_k}
\frac{p_\theta(z_{j-1}^{(m)}\mid z_j^{(m)})}
     {q(z_j^{(m)}\mid z_{j-1}^{(m)})},
\qquad
\overline R_k
=
\frac{1}{M}\sum_{m=1}^{M}R_{km},
\end{equation}
where $z_j^{(m)}$ is the state of candidate $m$ at step $j$.
In the work of the original path $(\Gamma,x)$, segment $k$ contributes $-\log R_{k1}$. We replace this contribution with $-\log\overline R_k$ for each segment. The resulting work used in BAR is
\begin{equation}
w_M
=
-\beta U(x)
+
\log\frac{Q(\Gamma\mid x)}{P_\theta(\Gamma,x)}
+
\sum_k \bigl(\log R_{k1}-\log\overline R_k\bigr).
\end{equation}
For $M=1$, this recovers the original work $w$. We use the same construction for forward and reverse paths. 
Using this formula preserves the partition function identity for any finite $M$ (\hyperref[sec:si-conditional-path-averaging]{Supplementary Section~\ref*{sec:si-conditional-path-averaging}}).

\subsection*{Learning amortized samplers from energy}

For capped peptides, each hierarchy level uses the same drift network across sequences. The network receives the atom identities and bond connectivity of the peptide being sampled. Training follows a fixed-point iteration based on Flow Sampling (Fig.~\ref{fig:peptides}a).\cite{havens2026flowsampling,blessing2026bridge} At each round, the current sampler generates complete molecular configurations. For each configuration $x$, we evaluate the potential energy and atomic forces using the energy function for its sequence. We store the configuration, its forces and the work $w(\Gamma,x)$ in a replay buffer.

One evaluation of the atomic forces supplies the force term for every hierarchy level. Let $x(y)$ denote the atomic positions reconstructed from the hierarchy coordinates $y$. Changing the coordinates $y_\ell$ at level $\ell$ can move several atoms, even when the other hierarchy coordinates are held fixed. The chain rule combines the forces on those atoms to give the reduced force on $y_\ell$,
\begin{equation}
g_\ell
=
-\nabla_{y_\ell}\bigl[\beta U(x(y))\bigr]
=
\left(\frac{\partial x}{\partial y_\ell}\right)^{\!\top}
\bigl[-\beta\nabla_x U(x(y))\bigr].
\end{equation}

To train the drift at level $\ell$, we pair a stored configuration with an independent Gaussian prior draw $\xi_\ell\sim\rho_\ell$. At a sampled time $t\in[0,1]$, the network receives the interpolated coordinates $z_t=(1-t)y_\ell+t\xi_\ell$ and the coarser structure $y_{<\ell}$. With $t=0$ at the molecular configuration and $t=1$ at the prior, the regression loss for this example is
\begin{equation}
L_\ell
=
\frac{1}{d_\ell}
\left\|
u_{\theta,\ell}(z_t,t;y_{<\ell})
-
\bigl(\xi_\ell-y_\ell-\lambda_\ell g_\ell\bigr)
\right\|^2,
\end{equation}
where $d_\ell$ is the number of coordinates at this level for the given peptide. We set $\lambda_\ell$ inversely proportional to the root-mean-square magnitude of $g_\ell$ among generated configurations in the buffer. The same coefficient sets the diffusion strength during sampling.

We weight the regression losses of generated configurations by $\exp(\alpha w)$, normalizing within each sequence and choosing $\alpha$ to maintain a prescribed effective sample size. A separate damping term penalizes the squared change in the predicted drift from the start of the round. The updated sampler then generates the next batch of configurations. For the shared sampler, reference MD configurations from 40 homopeptides within the 800 training sequences supply one quarter of each training batch and receive unit weight (\hyperref[sec:sampler]{Supplementary Section~\ref*{sec:sampler}}).

\subsection*{Network architectures}

For proteins and host-guest complexes, each hierarchy level uses an equivariant graph transformer, and the finer levels share an encoder of the coarse structure.\cite{tholke2022torchmd} For the energy-trained peptide sampler, each level uses a PaiNN-based network shared across peptide sequences and chain lengths.\cite{painn2021}

At each finer level, we reconstruct the Cartesian positions of the atoms being generated from their noisy diffusion coordinates. 
For an atom represented by a relative displacement, we multiply its diffusion coordinate by the length scale of the level and add the position of its parent atom. 
We then build the graph containing these atoms and the fixed atoms from coarser levels. Both sets of atoms participate in message passing. The network returns a Cartesian update for each atom being generated, which we then transfer back to obtain its update in diffusion coordinates.
For constrained X-H bonds, we express the update in the two coordinates that describe the bond direction. 
For bulk systems, a three-dimensional convolutional network with circular padding models the cell counts.

\subsection*{Molecular systems}

The four protein systems were the Trp-cage miniprotein TC10b, the wild-type B1 domain of protein G (GB1), the albumin-binding GA domain 1PRB and the designed three-helix bundle 2A3D. We used the Amber ff14SB force field with GBn2 implicit solvent and constrained X-H bonds for all four proteins.\cite{maier2015ff14sb,nguyen2013gbn2} We generated reference ensembles and also obtain the free energy reference by non-reversible parallel tempering (NRPT),\cite{syed2022non} varying the solvent dielectric across replicas for TC10b and 1PRB. For GB1, we ran NRPT across the alchemical states of the Y33G mutation in the protein and of the corresponding Tyr-to-Gly substitution in a capped tripeptide.

The host-guest systems comprised tetra-\textit{endo}-methyl octa-acid (TEMOA) from SAMPL6 and tetra-\textit{endo}-ethyl octa-acid (TEETOA) from SAMPL8, each paired with guests G1, G4 and G5 from its respective challenge.\cite{rizzi2018sampl6,amezcua2022sampl8} We used GAFF 1.8 for the TEMOA complexes and GAFF2 for the TEETOA complexes, both with AM1-BCC charges and GBn2 implicit solvent. We ran NRPT at $298.15\,\mathrm{K}$, varying the host-guest interaction from fully coupled to decoupled across replicas. For both the protein and host-guest calculations, we obtained reference free energy differences by applying the multistate Bennett acceptance ratio (MBAR) to samples from all states.\cite{shirts2008mbar}

The peptide systems contained one to four amino acid residues between acetyl and N-methylamide caps. The shared sampler was trained on 300 two-residue and 500 three-residue sequences. We modeled these peptides with Amber ff14SB and OBC2 implicit solvent, leaving X-H bonds flexible. We generated a reference ensemble for each evaluated sequence by molecular dynamics (MD) at $300\,\mathrm{K}$. For ablation, we compared hierarchy depths on alanine dipeptide (ACE-Ala-NME), a benchmark for Boltzmann sampling.\cite{midgley2022flow} We used Amber ff96 with OBC2 implicit solvent and generated the reference by MD at $300\,\mathrm{K}$.

The bulk CO$_2$ systems were periodic boxes containing 216 molecules for training and up to 1,728 molecules for size transfer. We used the rigid TraPPE model for CO$_2$.\cite{potoff2001trappe} We generated reference ensembles by constant-volume Langevin MD at $320\,\mathrm{K}$, covering six densities from $0.040$ to $0.400\,\mathrm{g\,cm^{-3}}$ at the training size. For each system size, we calculated reference compression free energy by applying MBAR across densities.

The metallic glass former was Cu$_{64}$Zr$_{64}$, an equiatomic alloy of 128 atoms. We modeled its atomic interactions with the MACE-MPA-0 interatomic potential.\cite{batatia2025foundation} We ran NRPT across temperatures to sample the glass at $700\,\mathrm{K}$ and the melt at $1{,}350$ and $2{,}000\,\mathrm{K}$. We obtained reference differences in reduced free energy by stepping-stone estimation along the ladder.\cite{xie2011steppingstone}

\section*{Acknowledgments}
JH acknowledges support from the University of Cambridge Harding Distinguished Postgraduate
Scholars Programme.

\clearpage
\include{SI/SI}

\bibliography{references.bib}

\end{document}

%% file: SI/SI.tex
\setcounter{section}{0}
\setcounter{equation}{0}
\setcounter{figure}{0}
\renewcommand{\thesection}{S\arabic{section}}
\renewcommand{\theequation}{S\arabic{equation}}
\renewcommand{\thefigure}{S\arabic{figure}}
\renewcommand*{\theHsection}{supp.\arabic{section}}
\renewcommand*{\theHfigure}{supp.\arabic{figure}}
\renewcommand*{\theHequation}{supp.\arabic{equation}}
\pdfbookmark[0]{Supplementary Information}{archimedes-supplement}
\begin{center}
{\Large\bfseries Supplementary Information}\par
{\normalsize\itshape Scalable and Accurate Generative Sampling of Molecular Equilibrium Ensembles}
\end{center}

\begingroup
\clubpenalty=10000
\widowpenalty=10000
\section{Diffusion Details}

\subsection{Molecular coordinates}

In proteins, we express each backbone or side-chain heavy atom as a displacement from the C$\alpha$ atom of its residue. For host-guest complexes, we select coarse atoms separately for the host and the guest. We start from the cyclic core of each molecular graph, or from its longest bond path if the molecule has no cycle. Each new coarse atom is chosen to maximize the shortest bond distance to those already selected. We assign each remaining heavy atom to its nearest coarse atom by bond distance. 
For the C$\alpha$ level, we remove overall translation and evaluate the prior on the space of centered coordinates.

For a constrained X-H bond, the hydrogen moves on a sphere around its bonded heavy atom. We describe its direction by two stereographic coordinates $u=(u_1,u_2)$ in a local frame defined by the heavy atoms. For bond length $r$, the unit direction $\mathbf n$ and the corresponding surface area element are
\begin{equation}
\mathbf n(u)=\frac{(2u_1,2u_2,1-\lVert u\rVert^2)}{1+\lVert u\rVert^2},
\qquad
\mathrm dA=\frac{4r^2}{(1+\lVert u\rVert^2)^2}\,\mathrm du_1\,\mathrm du_2 .
\end{equation}

\subsection{Gaussian transitions in continuous diffusion}

Consider one continuous level fitted to reference configurations. Let $z_0$ denote its clean coordinates and $\sigma$ its prior width. The coarser coordinates stay fixed. For two diffusion times $0\leq s<t\leq1$, set $a=(1-t)/(1-s)$. The forward transition is
\begin{equation}
q(z_t\mid z_s)=\mathcal N\!\left(a z_s,\,\sigma^2(t^2-a^2s^2)I\right).
\end{equation}
Starting at $s=0$ gives $z_t\mid z_0\sim\mathcal N((1-t)z_0,\,t^2\sigma^2I)$. At $t=1$, this is the Gaussian prior for every $z_0$.

To construct the reverse transition, first suppose the clean coordinates are known. Gaussian conditioning gives a mean $b z_t+c z_0$ and covariance $vI$ for $z_s$, where
\begin{equation}
 b=\frac{as^2}{t^2},\qquad c=1-s-b(1-t),\qquad v=\sigma^2s^2(1-ab).
\end{equation}
Given $z_t$, the time $t$ and the coarser structure, the network predicts the clean coordinates $\hat z_0$ and their uncertainty $C_\theta$. The learned reverse transition is
\begin{equation}
p_\theta(z_s\mid z_t)=\mathcal N\!\left(bz_t+c\hat z_0,\,vI+c^2C_\theta\right).
\label{eq:si-gaussian-reverse}
\end{equation}
For the protein and host-guest models, a scalar network output $\ell_{\theta,i}$ rescales the reference standard deviation for atom $i$. The resulting variance is
\[
d_i=B_i(t)\exp(2\ell_{\theta,i}),\qquad
B_i(t)=\frac{r_i^2\,t^2\sigma^2}{(1-t)^2r_i^2+t^2\sigma^2},
\]
where $r_i$ is the root-mean-square scale of the clean diffusion coordinates. The centered coarse level averages the scalar outputs over atoms and uses $C_\theta=dI$ on the centered subspace. At finer levels, $C_\theta$ is block diagonal in the diffusion coordinates, with one block per atom. For an isotropic block, $C_{\theta,i}=d_i I$, including $I_2$ for a constrained hydrogen. For the Cartesian backbone and side-chain levels of the protein models, the network outputs two scalars per atom. They give $d_{i,\parallel}$ and $d_{i,\perp}$ by the same formula, with
\[
C_{\theta,i}
=d_{i,\parallel}u_i u_i^\top+d_{i,\perp}(I-u_i u_i^\top),
\]
where $u_i$ is the unit vector from the parent atom to the noisy atom. The CO$_2$ position and orientation levels each use an isotropic block per molecule and predict the log variance multiplier directly, giving $d_i=B_i(t)\exp(\ell_{\theta,i})$.

We train $\hat z_0$ and $C_\theta$ jointly by maximizing the path-space evidence lower bound (ELBO) over reference configurations. On a grid $0=t_0<\cdots<t_K=1$, the ELBO for a clean configuration $z_0$ is
\[
\begin{aligned}
\mathcal L_{\mathrm{ELBO}}(z_0)
&=\mathbb E_{q(z_{t_1}\mid z_0)}
  \big[\log p_\theta(z_0\mid z_{t_1})\big]\\
&\quad-\sum_{j=2}^{K}
  \mathbb E_{q(z_{t_j}\mid z_0)}
  \!\left[D_{\mathrm{KL}}\!\left(
  q(z_{t_{j-1}}\mid z_{t_j},z_0)
  \,\big\|\,p_\theta(z_{t_{j-1}}\mid z_{t_j})
  \right)\right].
\end{aligned}
\]
The terminal KL term is zero because $q(z_{t_K}\mid z_0)$ equals the prior. We use more closely spaced diffusion times near $t=0$ to resolve the final denoising steps.

\subsection{Cell counts and positions in bulk systems}

The occupancy level uses masked discrete diffusion to generate one cell count at a time. Let $p_k$ be the categorical probability of the count generated at step $k$, given the counts already revealed. The probability of the reveal order cancels from the work because the noising process masks cells in the opposite order with the same probability.

There are $N!/\prod_s n_s!$ ways to assign $N$ labeled molecules to cells with counts $n_s$. We draw one of these assignments uniformly. In the forward direction, the cell labels $a$ are drawn from the probabilities $r(a\mid x)$ defined in Methods. The contribution of counts and assignments to the work is therefore
\begin{equation}
\log r(a\mid x)-\sum_k\log p_k+\log\!\left(\frac{N!}{\prod_s n_s!}\right).
\end{equation}
The occupancy CTMC localizes each molecule to the neighborhood of its assigned cell, so the position model learns short displacements from the cell center. The forward process uses the minimum-image displacement, and the work calculation assumes that the probability of generating a displacement beyond half the box length along any coordinate is negligible.

For a rigid CO$_2$ molecule, a separate level represents the molecular axis by the direction of a three-dimensional vector. An isotropic Gaussian prior gives a uniform axis direction.
The model represents each CO$_2$ orientation by a three-dimensional vector whose direction gives the molecular axis. For each MD configuration, we multiply the unit vector along the molecular axis by the length of an independent draw from the Gaussian prior. Normalizing the resulting vector recovers the original orientation.

\section{Free energy estimation}

\subsection{Forward and reverse work distributions}

We use the work convention of the Methods. The log importance weight is $-w_f$ for a forward path and $w_b$ for a reverse path. For $N_{\mathrm F}$ forward paths and $N_{\mathrm B}$ reverse paths, exponential averaging gives two estimates of $\log Z$,
\begin{equation}
\widehat{\log Z}_{\mathrm F}
=-\log\!\left(\frac{1}{N_{\mathrm F}}\sum_i e^{w_{f,i}}\right),
\qquad
\widehat{\log Z}_{\mathrm B}
=\log\!\left(\frac{1}{N_{\mathrm B}}\sum_j e^{w_{b,j}}\right).
\end{equation}
The forward-reverse gap measures the difference between these estimates. Work separation is $\min_i(-w_{f,i})-\max_j(w_{b,j})$. A positive value is the distance from the largest sampled reverse work to the smallest sampled forward work.

\emph{Why does pos-generation relaxation not help?}
A relaxation step appended to a generated configuration leaves its work unchanged if it satisfies detailed balance with the target. Let $K(x'\mid x)$ be the probability density for relaxing $x$ to $x'$. The change in the Boltzmann weight and the forward and reverse transition probabilities contribute a factor
\begin{equation}
\frac{e^{-\beta U(x')}K(x\mid x')}{e^{-\beta U(x)}K(x'\mid x)}=1
\label{eq:si-relaxation}
\end{equation}
to the importance weight. The equality follows directly from detailed balance.

\subsection{Conditional path averaging}

Fix the endpoints $z_a$ and $z_b$ of a diffusion segment. Averaging over the intermediate states removes the variation in the segment weight due to paths between these two endpoints. To see this, let $\eta$ denote the intermediate states. The product of forward transition densities along the segment is $q(\eta,z_b\mid z_a)$. The product in the reverse direction is $p_\theta(\eta,z_a\mid z_b)$. The forward Gaussian bridge has density $B(\eta)=q(\eta,z_b\mid z_a)/q(z_b\mid z_a)$. Under this bridge, the conditional expectation of the segment ratio $R$ defined in Methods is
\begin{equation}
\mathbb E_B[R]
=\int B(\eta)\frac{p_\theta(\eta,z_a\mid z_b)}{q(\eta,z_b\mid z_a)}\,\mathrm d\eta
=\frac{p_\theta(z_a\mid z_b)}{q(z_b\mid z_a)} .
\end{equation}
The last expression depends only on the endpoints. The Gaussian bridge lets us estimate this average while keeping the molecular configuration and its energy fixed.

\leavevmode\phantomsection\label{sec:si-conditional-path-averaging}%
In practice we retain the original trajectory and draw $M-1$ additional trajectories from the bridge. On a forward path, all $M$ candidates follow the bridge. On a reverse path, the original trajectory was generated by the learned reverse process. To obtain the density ratio for all the candidates, we can place the original trajectory at a uniformly chosen index without changing the average. Writing $\eta_m$ for candidate $m$, the ratio of the reverse and forward densities for the segment is
\begin{equation}
\frac{\displaystyle\frac{1}{M}\sum_{m=1}^{M}p_\theta(\eta_m,z_a\mid z_b)
                      \prod_{j\ne m} B(\eta_j)}
     {\displaystyle q(z_b\mid z_a)\prod_{j=1}^{M}B(\eta_j)}
=\frac{1}{M}\sum_{m=1}^{M}R(\eta_m).
\end{equation}
Thus the contribution to the work is the negative logarithm of the mean ratio. Across all segments, this gives the ratio of path densities used to calculate $w_M$ in Methods. The added bridge densities each integrate to one, so $\mathbb E[e^{w_M}]=Z$ over generative paths for any $M$. The same work $w_M$ can therefore be used in BAR.

\subsection{Protein free energy differences}

For TC10b, we evaluated the solvent free energy change three times, using $4{,}096$ paths per direction at each dielectric and pairing the endpoint estimates within each evaluation. In the comparison across training checkpoints at $\epsilon=78.5$, we split the paths into two groups of up to $2{,}048$ per direction. We pooled both groups for the work histograms.

In the hierarchy comparison, we did not fix X-H bond for the one-level model, while for both two- and ours four model, we use fixed X-H bond.

For 1PRB, each of the three paired solvent free energy evaluations used $1{,}024$ paths per direction at each dielectric, with $M=16$ trajectories per segment for conditional path averaging.

For GB1 Y33G, each protein endpoint used up to $4{,}096$ paths per direction with $M=64$ trajectories per segment. Each capped tripeptide endpoint used $1{,}024$ paths per direction. We subtracted the same tripeptide free energy difference from each of the three protein estimates.

\subsection{Relative binding free energies}

For TEMOA and TEETOA, we used $4{,}096$ paths per direction for each coupled and decoupled endpoint. We combined the endpoint estimates within each of three evaluations. The same flat-bottom restraint acts on the distance between the host and guest centers of mass in both endpoints for a given guest.

\leavevmode\phantomsection\label{sec:si-binding-offset}%
Let $\Delta F_i=F_i^{\mathrm{coupled}}-F_i^{\mathrm{decoupled}}$ for guest $i$. Its standard binding free energy is $\Delta G_i=\Delta F_i-k_{\mathrm B}T\log(V_i/V^\circ)$, where $V^\circ=1.66054\,\mathrm{nm}^3$ is the volume per molecule at $1\,\mathrm{M}$. For a flat-bottom restraint of radius $R_i$ and force constant $k_i$, the potential at separation $r$ is $U_i^{\mathrm{rest}}(r)=\tfrac12 k_i\max(0,r-R_i)^2$. Its effective volume is $V_i=4\pi\int_0^\infty r^2 e^{-\beta U_i^{\mathrm{rest}}(r)}\,\mathrm dr$. The relative binding free energy is therefore
\[
\Delta\Delta G_{ij}=\Delta F_i-\Delta F_j+k_{\mathrm B}T\log(V_j/V_i).
\]
All three TEMOA guests share the same restraint volume. For TEETOA we used $V_{\mathrm{G1}}=V_{\mathrm{G5}}=1.3654\,\mathrm{nm}^3$ and $V_{\mathrm{G4}}=0.8100\,\mathrm{nm}^3$. The offset is $-0.522\,k_{\mathrm B}T$ for G1 minus G4, $+0.522\,k_{\mathrm B}T$ for G4 minus G5, and zero for G1 minus G5. We compared these differences with the reported MBAR binding free energies using the same guest order.

\subsection{Free energy differences in bulk systems}

At each CO$_2$ density and box size, the energy histogram compares $512$ generated configurations with $8{,}192$ MD configurations. For free energy estimation, each of four paired BAR evaluations used $128$ paths per direction at each density. The work histograms pool the four seeds for better visualization. 
The CO$_2$ work is evaluated with one factor of the box volume $V$ removed for overall translation. 
We restore this factor by adding $\log V$ to $\log Z$ at each density. 
At each size, we report compression relative to the lowest evaluated density.

For Cu$_{64}$Zr$_{64}$, the energy histogram uses $512$ generated configurations per temperature. Again, each of four paired BAR evaluations used $128$ paths per direction at each temperature and the work histograms pool the four evaluations. 
We report $[\beta_T F(T)-\beta_{T_0}F(T_0)]/N$ relative to $T_0=2{,}000\,\mathrm K$, where $\beta_T=(k_{\mathrm B}T)^{-1}$. The stepping-stone reference combines partition function ratios between adjacent temperatures in the NRPT ladder. 
At $700\,\mathrm K$, 
we raised the predicted probabilities for each cell count of the CTMC to the power $1/0.9$ and normalized them to sum to one. 
Also at $700\,\mathrm{K}$, individual NRPT chains remained in distinct amorphous basins, so we pooled configurations across the training chains to calculate the free energy. 
We used these probabilities both to draw counts and to evaluate the work. At the two melt temperatures, we used the network probabilities directly.

\section{Training and evaluation}

\subsection{Learning amortized samplers from energy}
\label{sec:sampler}

Our training pipeline and object follows mostly the flow sampler.\cite{havens2026flowsampling}
Also following previous works, we incorporate the tempered importance weights into the training objective.
The tempering of the weight is to maintain a target effective sample size.\cite{cheng2026atlas,he2026digin} We normalize the weights separately for each peptide sequence and use a higher target early in training. These weights apply only to the regression loss. All configurations contribute equally to the damping term.

The reference MD simulations for the 40 homopeptides used constrained X-H bonds, whereas the peptide sampler uses flexible bonds.
We retain the heavy-atom coordinates from MD and periodically regenerate the hydrogen positions with the current sampler during training. 
These configurations receive unit regression weight. 
These MD reference data help to avoid mode missing.

For ACE-Ala-NME for the hierarchy ablation, we trained a one-level and a four-level sampler separately using molecular energies and forces. Both used Amber ff96 with OBC2 implicit solvent and the same flat-bottom restraint on alanine chirality.

\subsection{Evaluation of generated ensembles}

We calculated distribution overlap using normalized histograms with common bins. If $p_b$ and $q_b$ are the fractions of generated and reference configurations in bin $b$, then $\mathrm{OVL}=\sum_b\min(p_b,q_b)$. An overlap of one indicates identical histograms.

For the four protein ensembles, we compared $1{,}024$ generated configurations for TC10b and $512$ each for GB1 wild type 1PRB and 2A3D, all against $4{,}096$ MD configurations per protein. 
We aligned the C$\alpha$ atoms to the minimized reference structure before calculating C$\alpha$ RMSD. We calculated the radius of gyration from the heavy-atom coordinates. For the backbone dihedral distributions, we pooled $(\phi,\psi)$ pairs from residues for which both angles are defined.

For the shared peptide sampler, we drew $2{,}000$ configurations per sequence and computed the energy and backbone dihedral overlaps.
 For four-residue peptides, we evaluated the samples both as generated and after $5\,\mathrm{ps}$ of Langevin relaxation in the same force field.

For Cu$_{64}$Zr$_{64}$, the partial radial distribution function $g_{\mathrm{CuZr}}(r)$ compares the number of Cu-Zr pairs in a spherical shell at distance $r$ with the number expected for uniformly distributed atoms at the same density. We identified Cu-centered icosahedra in the reference configurations by common-neighbor analysis. The central Cu atom has 12 neighbors. For each of these 12 bonds, the two bonded atoms share five neighbors arranged in a ring.
\newpage
\section{Additional results and visualization}

\subsection{Hierarchy of guest host systems}

\begin{figure}[!htbp]
  \centering
  \includegraphics[width=\linewidth]{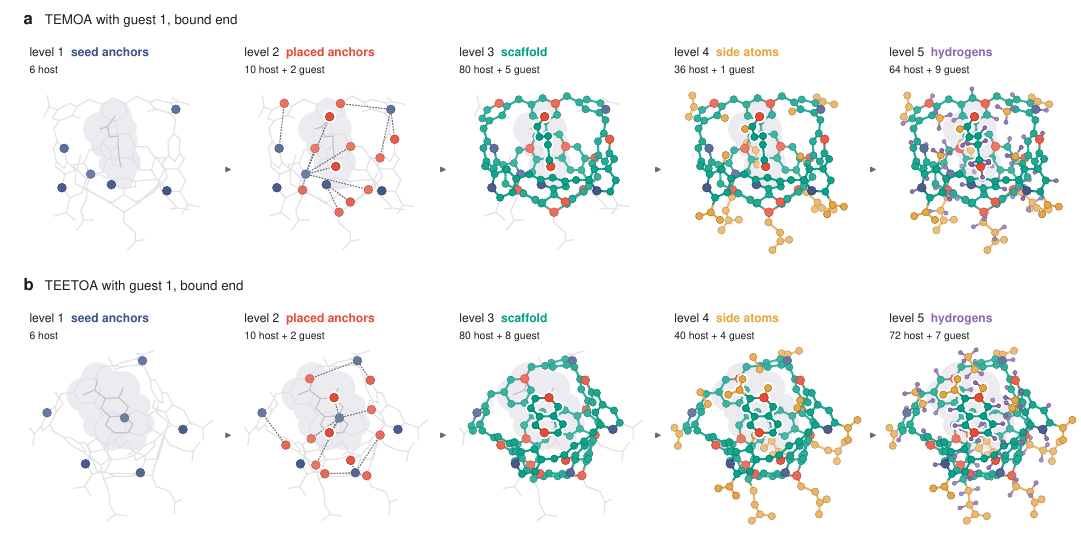}
  \caption{\textbf{Five-level generation of host-guest complexes.}
  \textbf{a,b}, Bound configurations of TEMOA with guest G1 ((\textit{E})-hex-2-enoate; \textbf{a}) and TEETOA with guest G1 (3-hydroxy-2-naphthoate; \textbf{b}). Each column adds one level.
  Level 1 places six host anchors. Level 2 adds the remaining ten host anchors and two guest anchors, establishing the guest's position relative to the host.
  Level 3 completes the molecular scaffolds, defined by the ring core of each host and the TEETOA guest, and by the longest heavy-atom chain of the TEMOA guest. Level 4 adds the remaining heavy atoms, and level 5 adds hydrogens at fixed bond lengths.
  Colours identify the levels, and grey shading marks the guest. Dashed lines connect each level-2 anchor to the level-1 anchor used to place it. Numbers below each panel give the numbers of host and guest atoms added at that level.}
  \label{fig:hg-levels}
\end{figure}

\newpage
\subsection{Conditional path averaging on GB1}

At a fixed number of network evaluations, allocating some evaluations to bridge trajectories narrowed the forward-reverse gap on GB1 (Supplementary Fig.~\ref{fig:path-averaging}a,b). For the work histograms, we used $4{,}096$ paths per direction for each protein and centered the work on its BAR estimate of $\log Z$ at $M=64$. For both wild type and Y33G, the sampled work ranges were separated with one trajectory per segment and overlapped with four (Supplementary Fig.~\ref{fig:path-averaging}c).

\begin{figure}[!htbp]
\centering
\includegraphics[width=\linewidth]{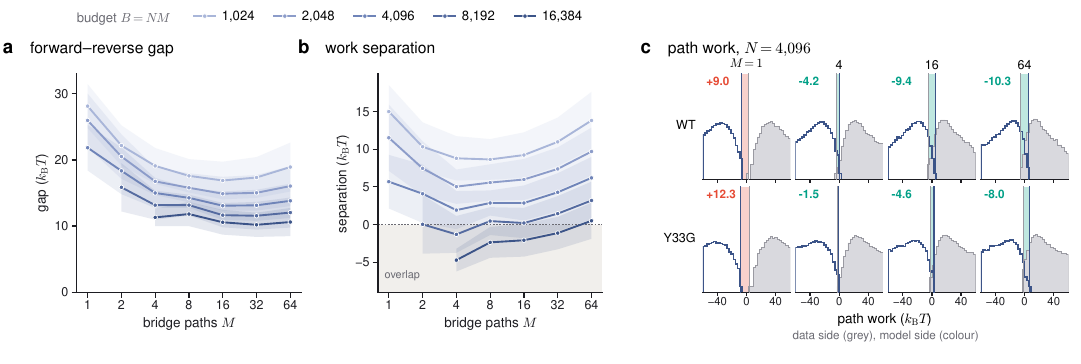}
\caption{\textbf{Conditional path averaging on GB1.}
\textbf{a,b}, Forward-reverse gap (\textbf{a}) and work separation (\textbf{b}), averaged over GB1 wild type and Y33G. Each curve holds $NM$ fixed, where $N$ is the number of paths per direction and $M$ is the number of trajectories per segment. Lines show means across three evaluations. Bands show the standard deviation of individual runs.
\textbf{c}, Forward ($-w_f$, grey) and reverse ($w_b$, blue) work distributions from one evaluation at fixed $N$. Numbers give the work separation in $k_{\mathrm B}T$. Green indicates overlap of the sampled ranges and red indicates a gap.}
\label{fig:path-averaging}
\end{figure}

\newpage
\subsection{Energy and work distributions for bulk systems}

\begin{figure}[!htbp]
\centering
\includegraphics[width=\linewidth]{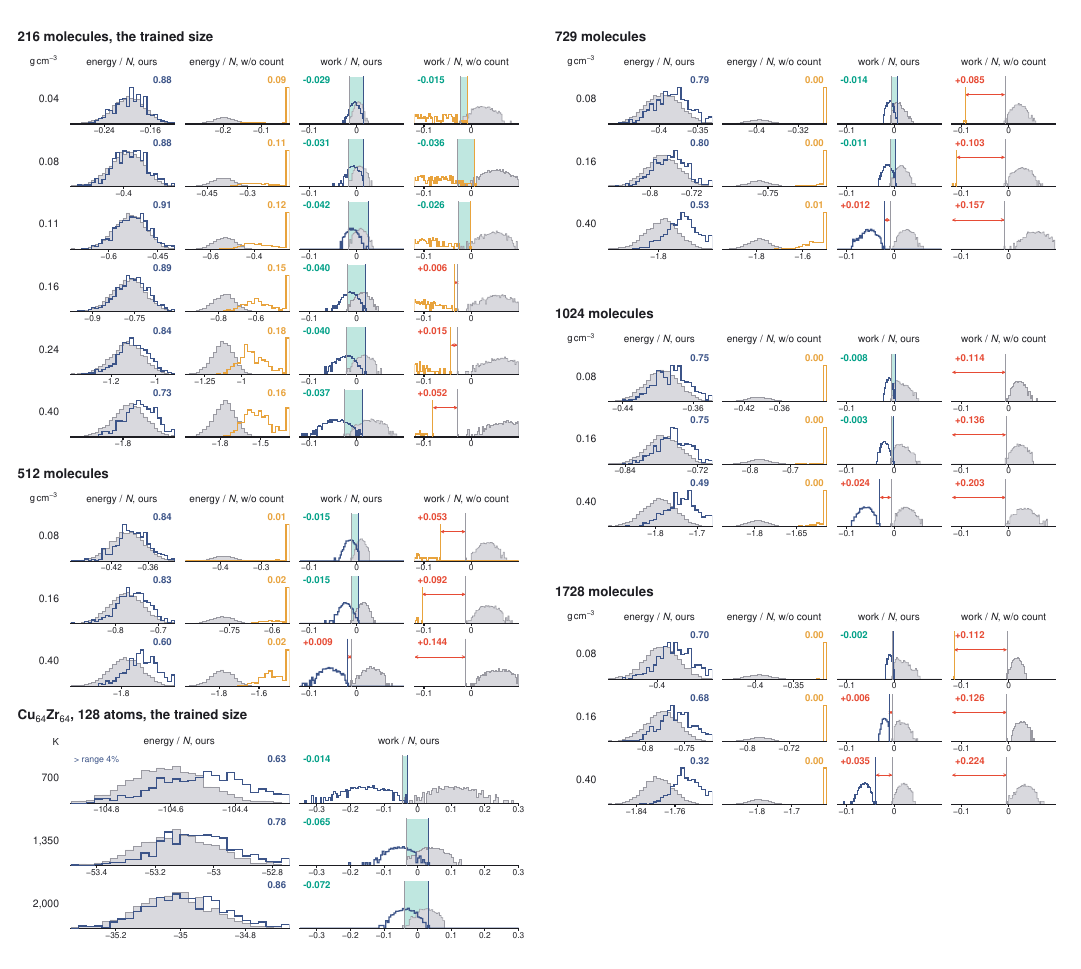}
\caption{\textbf{Energy and work distributions for bulk CO$_2$ and Cu$_{64}$Zr$_{64}$.}
Each CO$_2$ block shows a box size, with one row per density. The Cu-Zr block shows one row per temperature. Energy panels compare the reduced potential energy per molecule or atom, $\beta U/N$, for \name{} (blue) and the CO$_2$ model without cell occupancies (orange) with the MD or NRPT reference (grey). Numbers give distribution overlap. Percentages indicate samples above the plotted range.
Work panels show $-w_f$ for forward paths (grey) and $w_b$ for reverse paths (model color). Both are centered on the forward estimate of $\log Z$ and divided by $N$. Numbers give the work separation per molecule or atom, computed from all sampled paths. Green indicates overlap of the sampled ranges and red indicates a gap.}
\label{fig:co2-cells}
\end{figure}
\clearpage
\endgroup